# Strain engineering of ultrafast magnetism in the room-temperature vdW ferromagnet $Fe_3GaTe_2$

*Fan Fei*[1,*], *Yangchen He*[1,*], *Wuzhang Fang*[1,*], *Jessica Kienbaum*[1], *Jacopo Simoni*[1], *Robert Boyd*[1], *Yuan Ping*[1,2,3,†], *Daniel A. Rhodes*[1,2,†], *Jun Xiao*[1,2,4,†]

[1] Department of Materials Science and Engineering, University of Wisconsin-Madison, Madison, Wisconsin 53706, USA

[2] Department of Physics, University of Wisconsin-Madison, Madison, Wisconsin 53706, USA

[3] Department of Chemistry, University of Wisconsin-Madison, Madison, Wisconsin 53706, USA

[4] Department of Electrical and Computer Engineering, University of Wisconsin-Madison, Madison, Wisconsin 53706, USA

* These authors contributed equally.

† Corresponding author(s). E-mail(s): yping3@wisc.edu, darhodes@wisc.edu, jun.xiao@wisc.edu

**Abstract**

Controlling ultrafast magnetic dynamics are critical to understanding nonequilibrium spin interactions and advancing high-speed spintronics. However, a lack of efficient in situ tuning strategies leaves most ultrafast studies largely dependent on the intrinsic properties of the individual materials. Here we demonstrate continuous strain tuning of both the equilibrium magnetic response and ultrafast demagnetization dynamics in the room-temperature van der Waals ferromagnet $Fe_3GaTe_2$. Applying up to 4.2% uniaxial tensile strain increases the coercive field from nearly zero to 100 Oe, consistent with an enhancement of the effective perpendicular magnetic anisotropy. Time-resolved magneto-optical Kerr effect measurements further reveal strain-accelerated ultrafast demagnetization, with 1.2% tensile strain reducing the characteristic demagnetization time by approximately 20%. Remarkably, strain accesses an accelerated demagnetization regime that cannot be reached simply by increasing pump fluence in the unstrained sample. Combined with first-principles calculations, our results resolve that the applied strain modifies the spin–lattice energy transfer, leading to the observed accelerated demagnetization. These findings establish mechanical strain as an effective route for on-demand control of ultrafast magnetic dynamics while reducing the required optical energy by reconfiguring the magnetic energy landscape and associated spin-relaxation pathways.

## Introduction

Ultrafast magnetic dynamics reveal how angular momentum and energy are redistributed within a magnetic system after transient excitation, thereby defining the efficiency and fundamental timescales of high-speed spintronic devices. To date, ultrafast magnetization dynamics driven by femtosecond optical excitation and picosecond electrical pulses have been demonstrated in a variety of magnetic systems[1–7], highlighting a viable route toward beyond-GHz magnetic operation. Achieving in situ control over these dynamics is essential for moving beyond passive observation toward actively programmable magnetic responses on ultrafast timescales. Despite this promise, practical strategies for continuously tuning ultrafast magnetic dynamics remain limited[8–10]. While modifying the underlying magnetic energy landscape, including exchange interactions and magnetic anisotropy, can effectively alter spin relaxation, commonly employed control parameters such as large magnetic fields or high pump fluences are not well suited for energy-efficient device operation[5,11,12]. This motivates the search for mechanically accessible materials platforms in which the magnetic energy landscape can be continuously reconfigured under modest external perturbations.

Two-dimensional van der Waals (2D vdW) magnets are particularly attractive in this context because their mechanical flexibility and strong sensitivity to lattice deformation enable substantial and continuous tuning of magnetic properties[13–23]. For example, strain can continuously modify perpendicular magnetic anisotropy (PMA) through magnetoelastic coupling, a key parameter governing the stability of out-of-plane magnetization[13,14], the energy barrier for spin-orbit torque switching in perpendicularly magnetized devices[24,25], and the dimensions of chiral spin textures such as magnetic skyrmions[26,27]. Beyond reshaping the equilibrium magnetic state, lattice deformation may also alter the microscopic pathways governing nonequilibrium spin relaxation. Because ultrafast magnetic dynamics arise from energy and angular momentum exchange among electronic, spin, and lattice subsystems, strain-induced changes in the magnetic energy landscape are expected to affect spin-relaxation timescales. Unlike increasing the optical pump fluence, mechanical strain may provide an independent route to tune the intrinsic relaxation pathway while reducing the optical energy required to access new dynamics. However, direct experimental investigations of strain-controlled ultrafast magnetic dynamics in vdW magnets remain scarce.

Here, we show that uniaxial tensile strain tunes both the equilibrium magnetic response and nonequilibrium spin dynamics of the room-temperature van der Waals ferromagnet $Fe_3GaTe_2$. $Fe_3GaTe_2$ provides a compelling platform for investigating strain-controlled ultrafast magnetism because it combines above-room-temperature ferromagnetism, strong perpendicular magnetic anisotropy, and a pronounced magneto-optical response[28,29]. Reported room-temperature skyrmion phases in $Fe_3GaTe_2$ further underscore the relevance of this materials family for topological spintronics and memory applications[30–34]. Reflective magnetic circular dichroism (RMCD) measurements reveal a pronounced increase in coercive field under tensile strain, consistent with an enhancement of the effective perpendicular magnetic anisotropy and supported by first-principles calculations of the magnetic anisotropy energy. Time-resolved magneto-optical Kerr effect (TR-MOKE) measurements further demonstrate that tensile strain accelerates the intrinsic picosecond demagnetization dynamics. Notably, strain enables a low-fluence accelerated relaxation regime that cannot be reproduced by simply increasing the excitation fluence in the unstrained sample. Combined with first-principles calculations, our results indicate that lattice deformation modifies the spin-lattice energy-transfer pathway. These findings establish mechanical strain as an effective control parameter for engineering both static magnetic properties and ultrafast spin relaxation in vdW magnets.

**Results**

$Fe_3GaTe_2$ crystallizes in centrosymmetric layered hexagonal structure with space group: $P6_3/mmc$, in which the Fe magnetic moments align ferromagnetically along the out-of-plane direction (**Figure 1a**). Moreover, recent studies suggest that finite Dzyaloshinskii–Moriya interaction may arise from inversion-symmetry broken interfacial effects or local Fe-site disorder/intercalation, resulting in robust skyrmion textures[31,32,34–36]. Single-crystalline $Fe_3GaTe_2$ samples were synthesized by chemical vapor transport method, whose stoichiometry was further confirmed by energy-dispersive X-ray spectroscopy (EDS) analysis (see **Methods** and Supplementary Fig. S2 for details). Bulk magnetization measurements under zero-field-cooled (ZFC) and field-cooled (FC) conditions reveal a ferromagnetic transition at approximately 366 K, confirming above-room-temperature magnetic order (**Figure 1b** and Supplementary Fig. S1). To characterize the room-temperature magnetic field dependent response of exfoliated $Fe_3GaTe_2$ flakes, we performed RMCD measurements at 298 K. As shown in **Figure 1c**, flakes with thicknesses of 33, 140, and 268 nm exhibit a pronounced thickness dependence of magnetic hysteresis (see Supplementary Fig. S3 for atomic force microscopy characterizations). With increasing thickness, the hysteresis loops become softer and the saturation field increases, consistent with an enhanced role of multidomain formation in thicker samples. This effect reduces the net remanent RMCD signal through domain averaging and requires a larger field to fully align magnetization[15,37]. Unless otherwise specified, the subsequent measurements were performed on flakes with a thickness of approximately 140 nm. This intermediate thickness provides a favorable balance between mechanical strain tunability, a sufficiently soft magnetic response for resolving strain-induced hardening, and a robust magneto-optical signal. These observations highlight $Fe_3GaTe_2$ as a versatile platform for investigating strain-tunable magnetic states at room temperature.

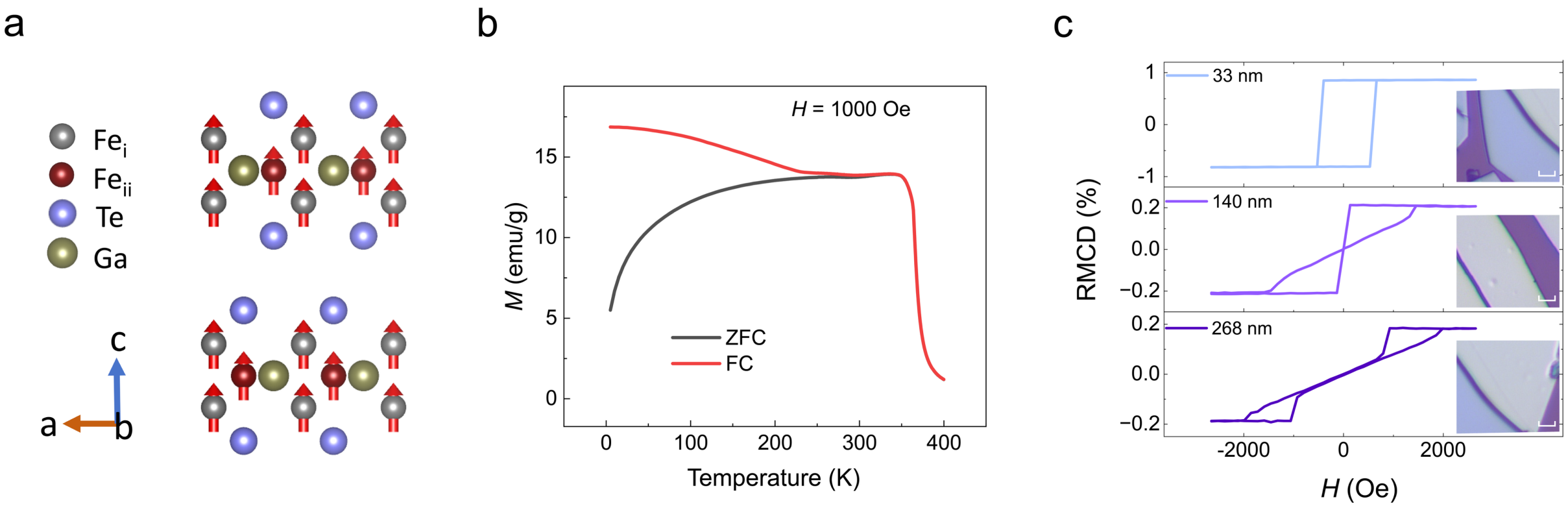

**Figure 1:** Crystal structure and room-temperature magnetic properties of $Fe_3GaTe_2$. **a,** Schematic crystal structure of $Fe_3GaTe_2$. Red arrows indicate the out-of-plane ferromagnetic alignment of the Fe magnetic moments. **b,** Temperature-dependent magnetization measured by SQUID magnetometry under zero-field-cooled (ZFC) and field-cooled (FC) conditions, showing a ferromagnetic transition near 366 K. **c,** Room-temperature RMCD hysteresis loops measured from exfoliated $Fe_3GaTe_2$ flakes with thicknesses of 33,

140, and 268 nm. The hysteresis loops exhibit a pronounced thickness dependence. Insets show optical microscope images of the corresponding flakes. Scale bars, 5 μm.

To investigate the strain dependence of the static magnetic responses, we fabricated a $Fe_3GaTe_2$/polymer/Si device mounted on a piezoelectric strain cell (**Figure 2a**). The $Fe_3GaTe_2$ flake was transferred onto a gapped Si substrate (initial gap distance ~5 μm) coated with a polycaprolactone (PCL) polymer layer, which mechanically anchors the flake across the gap region[38]. Increasing the piezoelectric displacement widens the gap and applies uniaxial tensile strain to the suspended $Fe_3GaTe_2$ flake. The strain was calibrated using a CrSBr flake placed on the same chip, based on its established strain-dependent phonon frequency[18] (see Supplementary Section S4 and **Methods** for more details). The strain-dependent evolution of the RMCD hysteresis is summarized in the two-dimensional RMCD map in **Figure 2b**, which clearly shows a strain-induced modification of the magnetic switching behavior. With increasing tensile strain, the hysteresis loop becomes more square, accompanied by a pronounced enhancement of the remanent RMCD amplitude and an increase in the saturated RMCD amplitude at high field. These observations indicate that tensile strain stabilizes the out-of-plane remanent magnetic state while also modifying the overall magneto-optical response. This trend is further illustrated by representative hysteresis loops in **Figure 2c**: compared with the unstrained state, the loop measured at 4% tensile strain exhibits a larger coercive field, enhanced remanent RMCD amplitude, and a larger saturated RMCD amplitude. The maximum accessible strain in the device was approximately 4.2%, beyond which further increasing the piezoelectric voltage caused partial strain relaxation (see more discussion and results in Supplementary Fig. S6).

To quantify these trends, we extracted the saturated RMCD amplitude (**Figure 2d**) and coercive field $H_c$ (**Figure 2e**) as functions of strain. The coercive field was defined as $H_c$ =

$(H_{\mathrm{up}} - H_{\mathrm{down}})/2$, where $H_{\mathrm{up}}$ and $H_{\mathrm{down}}$ denote the zero-crossing fields of the RMCD hysteresis loop during the up-sweep and down-sweep branches, respectively. The absolute RMCD signal measured at 1800 Oe increases approximately linearly from 0.15% at zero strain to 0.23% at 4.2% tensile strain. Over the same strain range, $H_c$ increases from nearly 0 Oe to approximately 100 Oe, with a tendency toward saturation at higher strain. The strain-induced increase in coercivity is consistent with an enhancement of the effective perpendicular magnetic anisotropy, although changes in domain nucleation and pinning may also contribute to the hysteresis evolution. To further assess the microscopic origin of this behavior, we performed first-principles calculations of the magnetic anisotropy energy under uniaxial tensile strain (**Figure 2f** and Supplementary Section S11). The calculations reveal an increase in the out-of-plane magnetic anisotropy energy with increasing strain, consistent with the experimentally observed hardening of the magnetic response. Together, these results establish uniaxial tensile strain as an effective in situ parameter for continuously tuning the static magnetic energy landscape of $Fe_3GaTe_2$ at room temperature.

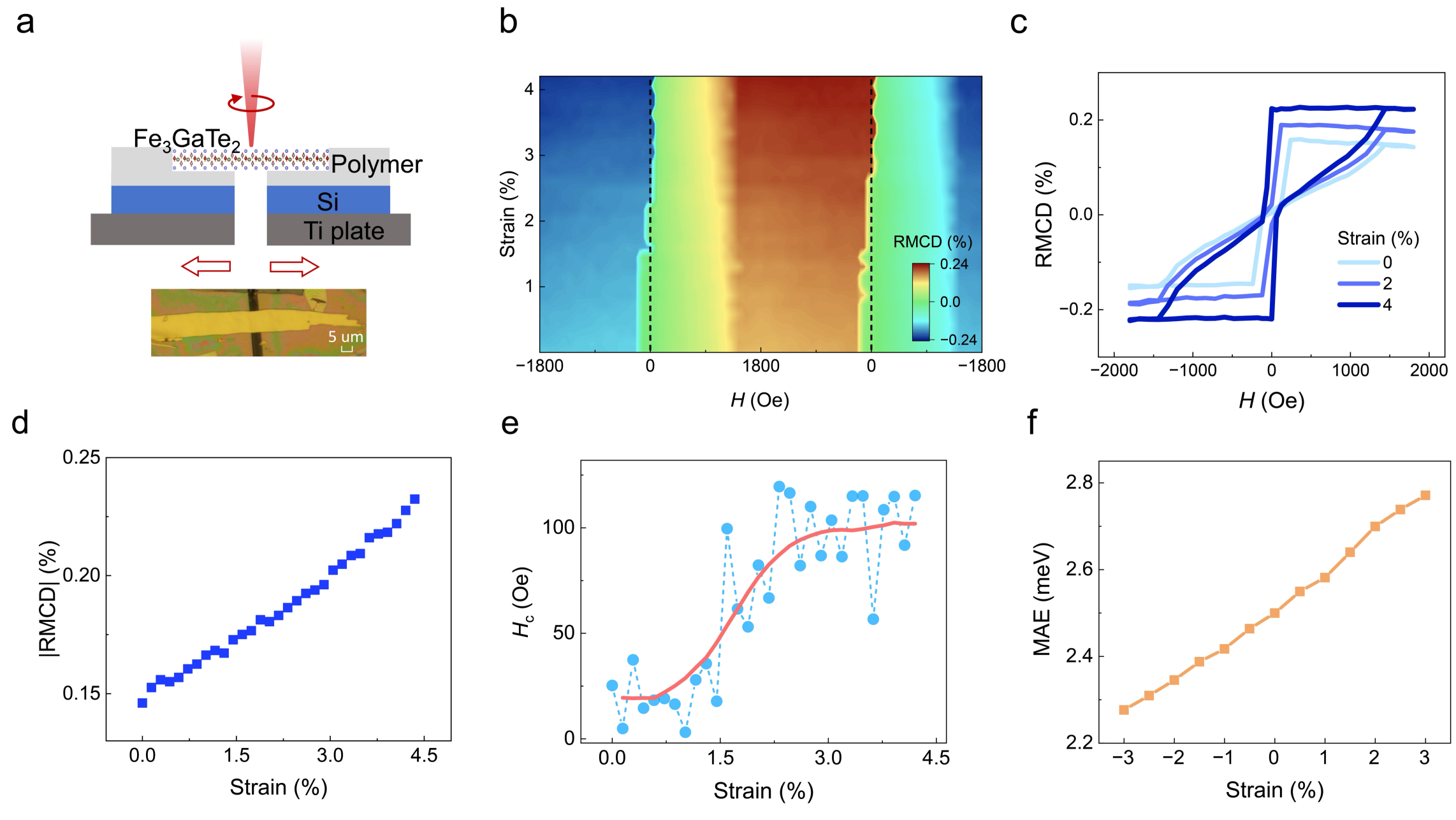

**Figure 2:** Strain tuning of the static magnetic response in $Fe_3GaTe_2$. **a,** Schematic illustration and optical microscope image of the strain device. An exfoliated $Fe_3GaTe_2$ flake is transferred onto a polycaprolactone (PCL)-coated gapped Si substrate mounted on a Ti plate. Piezoelectric displacement widens the gap and applies uniaxial tensile strain to the suspended $Fe_3GaTe_2$ region. The PCL thickness is exaggerated for clarity. The magnetic response is probed by RMCD, as indicated by the red beam. **b,** Two-dimensional map of the RMCD signal as a function of magnetic field and tensile strain, showing the progressive enhancement of the remanent signal and coercive field. **c,** Representative room-temperature RMCD hysteresis loops measured at selected tensile strains. **d,** Absolute RMCD signal measured at 1800 Oe as a function of tensile strain. **e,** Coercive field $H_c$, extracted from the RMCD hysteresis loops as a function of tensile strain. The solid pink line is a guide to the eye. **f,** First-principles calculations of the magnetic anisotropy energy (MAE) under uniaxial tensile strain, showing an enhancement of the perpendicular magnetic anisotropy with increasing strain.

Having established strain control of the static magnetic energy landscape, we next characterize the intrinsic ultrafast demagnetization dynamics of $Fe_3GaTe_2$ in the absence of applied strain. Time-resolved magneto-optical Kerr effect measurements were performed at room temperature using a 400 nm pump and an 800 nm probe in a polar geometry (see **Methods**). To minimize cumulative laser heating[11,39,40], the measurements were conducted at a repetition rate of 1 kHz. We first examined the magnetic-field dependence of the transient response at a pump fluence of 1 mJ/cm². As shown in **Figure 3a**, the demagnetization signal is weak near zero field and increases progressively with the applied out-of-plane magnetic field before saturating at approximately 1600 Oe. This trend is consistent with the gradual alignment of magnetic domains and the corresponding increase in the net magneto-optical response. Importantly, the overall temporal profile remains nearly unchanged as the magnetic field increases.

To quantify the dynamics, the transient response was fitted using a phenomenological two-component demagnetization model:

$$\Delta\varepsilon(t) = y_0 + A_1\left(1 - e^{-\frac{t}{\tau_{M,1}}}\right) + A_2\left(1 - e^{-\frac{t}{\tau_{M,2}}}\right). \quad (1)$$

where $A_1$ and $A_2$ denote the amplitudes of the fast and slow demagnetization components, respectively. $\tau_{M,1}$ and $\tau_{M,2}$ are the corresponding characteristic timescales. At a pump fluence of 1 mJ/cm², the response is dominated by the slow component, and the fast component cannot be reliably resolved. We therefore fit the low-fluence data using only the slow relaxation channel. A common fitting window of 0−70 ps was used throughout the manuscript to enable direct comparison among datasets. The extracted field-dependent fitting parameters are summarized in **Figure 3b**. The slow-component amplitude $A_2$ increases with magnetic field and closely follows the static RMCD response, further supporting its origin in the net magnetization of the aligned domains. In contrast, the slow relaxation time $\tau_{M,2}$ remains nearly field independent, with a value of about 16.9 ps. Given the field-dependent magnetic domain topology transitions in $Fe_3GaTe_2$[30,33], this field invariance indicates that the slow picosecond timescale is not primarily governed by changes in the macroscopic domain configuration in our 140-nm-thick flakes, but instead reflects an intrinsic relaxation pathway of $Fe_3GaTe_2$. Indeed, large-scale domain-related responses such as domain rearrangement, or skyrmion nucleation are generally slower and often occur in the sub-nanosecond-to-nanosecond regime[41,42]. Moreover, we did not observe the coherent skyrmion excitations in $Fe_3GaTe_2$ in the long-delay-time scans (Supplementary Fig. S9), which may be attributed to a large damping ratio caused by material defects[35].

We next investigated the fluence-dependent demagnetization dynamics by varying the pump fluence from 1 to 15 mJ/cm² under an out-of-plane magnetic field of 1800 Oe (**Figure 3c** and Supplementary Fig. S10 for full data). At low fluence, the transient response is dominated by the slow demagnetization channel. When the fluence exceeds 3.8 mJ/cm², an additional fast component occurring on a few-picosecond timescale becomes clearly resolvable. Furthermore, both the slow and fast relaxation times exhibit a strong fluence dependence. To quantify this behavior, we fit the data using Equation (1). To avoid overparameterization, we exclude the fast process for fluences below 3.8 mJ/cm². The extracted amplitudes and relaxation times are summarized in **Figure 3d**. We find that $A_2$ initially increases with pump fluence, reaches a maximum near 5 mJ/cm² and then gradually decreases. The corresponding slow relaxation time $\tau_{M,2}$ exhibits a similar nonmonotonic fluence dependence. By contrast, the fast-component amplitude $A_1$, increases continuously with fluence, while the fast relaxation time $\tau_{M,1}$ starts to decrease under strong excitation.

The monotonic increase of $A_1$ is primarily due to the strengthening of laser-induced hot electron–spin interactions with increasing pump fluence[43,44]. Accordingly, the fast relaxation time $\tau_{M,1}$ reflects spin relaxation mediated by excited-electron exchange interactions, which becomes faster at higher fluence. In contrast, the slow demagnetization channel is mainly governed by spin–lattice coupling[45,46]. Its amplitude $A_2$ initially increases with fluence as the overall demagnetization becomes stronger, but decreases once the fast demagnetization channel becomes dominant. The nonmonotonic evolution of the slow demagnetization time $\tau_{M,2}$ arises from critical slowing down near $T_c$[47–50]. When the transient spin temperature approaches $T_c$, $Fe_3GaTe_2$ will experience strong spin fluctuation and critical slowing down, a common phenomenon in various 2D magnets such as $FePS_3$ and $Fe_3GeTe_2$[39,51,52]. This effect contributes to the initial increase of $\tau_{M,2}$ with fluence. At higher fluence, this critical slowing down effect begins to saturate, and the increased lattice temperature and phonon population can enhance spin–lattice relaxation. This mechanism leads to the subsequent decrease of $\tau_{M,2}$, consistent with our previous work on the temperature-dependent spin–phonon relaxation rate in bcc Fe[53].

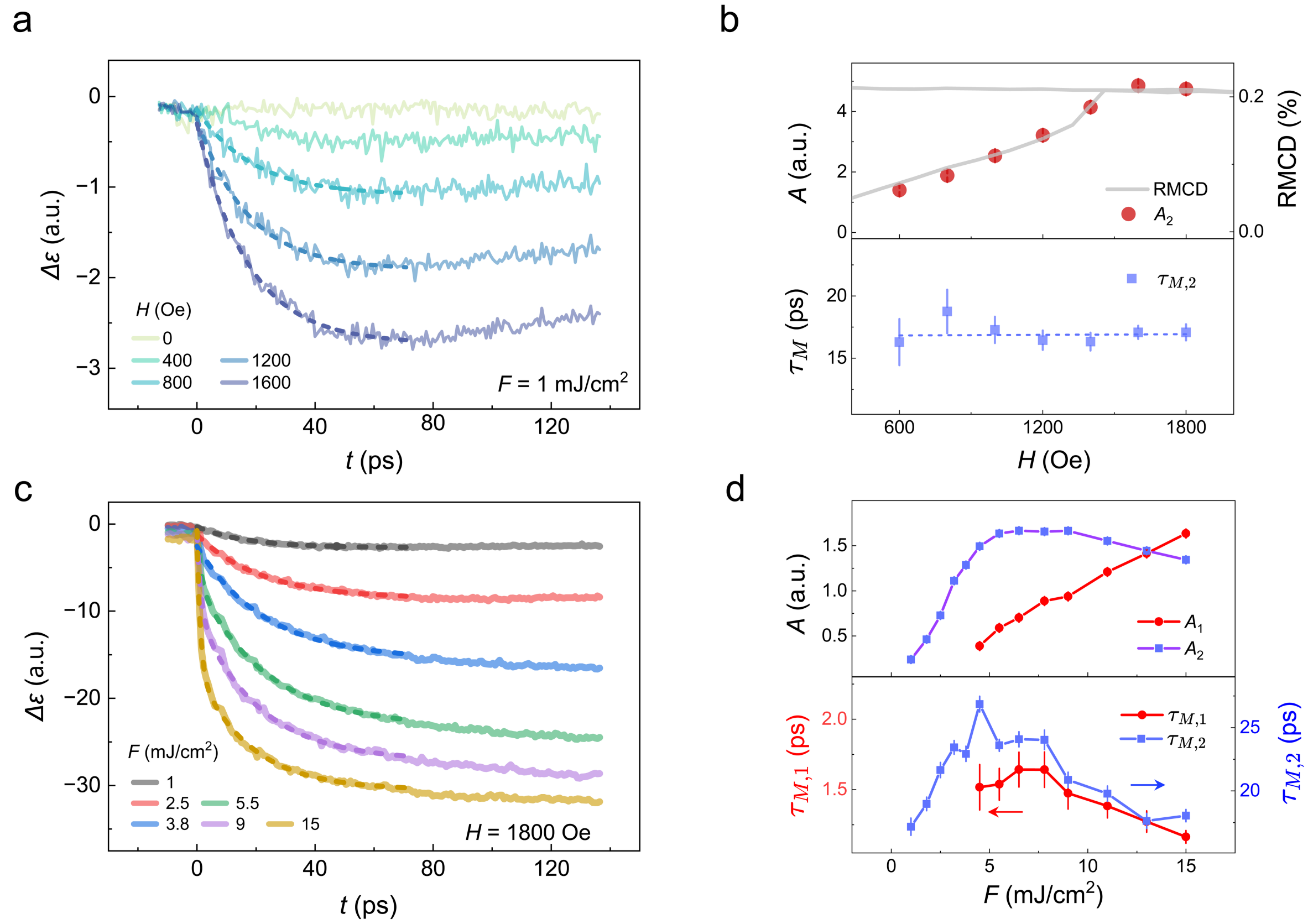

**Figure 3:** Ultrafast demagnetization in unstrained $Fe_3GaTe_2$. **a,** Transient Kerr ellipticity change $\Delta\varepsilon$ measured at room temperature under out-of-plane magnetic fields ranging from 0 to 1600 Oe. The pump fluence is 1 mJ/cm$^2$. Solid lines represent experimental data, and dashed lines denote fits. **b,** Field dependence of the slow-component amplitude $A_2$ and time constant $\tau_{M,2}$, extracted from the data in **a**. $A_2$ (red dots) scales with the static RMCD signal (red curve), whereas $\tau_{M,2}$ is nearly independent of field (blue dashed line). **c,** Fluence-dependent transient Kerr ellipticity change measured under an out-of-plane magnetic field of 1800 Oe. Solid lines represent experimental data, and dashed lines are fitted using Equation (1). Increasing fluence enhances the demagnetization amplitude and reveals a clear two-step relaxation behavior. **d,** Fluence dependence of the extracted amplitudes ($A_1$, $A_2$) and relaxation times ($\tau_{M,1}$, $\tau_{M,2}$) for the fast and slow components, respectively.

Having established the intrinsic ultrafast demagnetization pathways of unstrained $Fe_3GaTe_2$, we next investigate whether these dynamics can be continuously tuned by in situ mechanical strain. **Figure 4a** illustrates the strain-dependent TR-MOKE measurement geometry. The straining configuration has been introduced in **Figure 2a** and described in the **Methods**. Here, a larger gap increases the effective strained region from ~5 to ~20 μm to match the probe spot size (~10 μm), but at the cost of a reduced maximum strain. The strain is recalibrated using a nearby CrSBr flake on the same chip (Supplementary Fig. S5). To probe the strain response close to the weak-excitation limit, all measurements were performed at a pump fluence of 1 mJ/cm$^2$ under an out-of-plane magnetic field of 1800 Oe (See Supplementary Fig. S11). Representative transient Kerr ellipticity traces measured at selected tensile strains are shown in **Figure 4b**. Two systematic trends emerge as the strain increases: the demagnetization amplitude becomes larger, and the transient response becomes faster. The latter effect is evident from the progressively steeper decay and the earlier arrival of the characteristic (1/e) demagnetization point. The extracted parameters are summarized in **Figure 4c**. The demagnetization amplitude increases approximately linearly with tensile strain, while the slow demagnetization time $\tau_{M,2}$ decreases by nearly 20%, from $\sim$17 ps at zero strain to $\sim$14 ps at 1.2% strain. Notably, this strain-accelerated demagnetization cannot be accessed by increasing the pump fluence in the unstrained sample, as shown in **Figure 3d**. Although higher fluence strengthens the initial fast demagnetization channel, the slow channel, which determines when the maximum demagnetization is reached, remains longer than 17 ps at high pumping fluence. This comparison indicates that strain does not simply mimic stronger optical excitation, but instead provides an independent means of tuning the intrinsic picosecond spin relaxation. We note that further increasing the applied strain does not lead to a stable higher-strain state because the strain partially relaxes (Supplementary Fig. S12). The smaller achievable effective strain in this large-gap device likely arises from two factors. First, for flakes of similar overall length, increasing the gap size reduces the anchored region, weakening the mechanical constraint. Second, a wider suspended region increases the probability of mechanical failure caused by pre-existing defects and local strain hot spots.

We now discuss the microscopic origin of strain dependence. As shown in **Figure 2d**, the static RMCD intensity increases by nearly 50% under 4% strain. In the present ultrafast measurements, the demagnetization amplitude increases by ∼18% at 1.2% strain, which is broadly consistent with static RMCD enhancement. This agreement suggests that the strain-induced increase in demagnetization amplitude mainly reflects the enhancement of the static magneto-optical signal. In contrast, the strain dependence of the relaxation time is nontrivial and directly reflects a change in the underlying spin relaxation pathway. Combined with the strain-enhanced magnetic anisotropy discussed above, the observed reduction in $\tau_{M,2}$ strongly suggests a strain-enhanced spin-lattice coupling. Motivated by this interpretation, we further perform first-principles calculations of the electron-phonon relaxation rate under strain, as shown in **Figure 4d**. The electron-phonon relaxation rate for spin-up and spin-down channel is calculated using the equation[54]:

$$G_\sigma(T) = \frac{\pi\hbar k_B \lambda\langle\omega^2\rangle}{g_\sigma(\epsilon_F)} \int_{-\infty}^{\infty} g_\sigma^2(\epsilon)\left(-\frac{\partial f(T)}{\partial\epsilon}\right) d\epsilon,$$

where $g_\sigma(\epsilon)$ is the spin-resolved electronic density of states as shown in Supplementary **Figure S16,** $f(\epsilon, T)$ is the Fermi–Dirac distribution at electron temperature $T$, and $\lambda\langle\omega^2\rangle$ characterizes the strength of electron–phonon energy relaxation. To make the connection with the experimentally observed strain-enhanced demagnetization more transparent, we plot the normalized electron-phonon relaxation rate $G/G_0$ as a function of electron temperature and strain as shown in **Figure 4d**. We take the saturation value for the rate as a function of temperature. In this context, the electron temperature is used as an effective parameter that represents the laser-induced electronic excitation and associated broadening of the electronic occupations around the Fermi level. It should therefore not be interpreted as the actual thermodynamic temperature of the electronic subsystem. We observe that under tensile strain, the relaxation rate of the spin-down channel changes more efficiently than that of the spin-up channel, indicating that strain modifies the imbalance between the two spin-resolved relaxation pathways. The spin-up relaxation rate is nearly insensitive to strain, whereas the spin-down relaxation rate exhibits a stronger strain dependence that increases by about 20% from 0% to 3% tensile strain, which is in the same order of magnitude change ($\tau_{M,2}$ decreases by ~18% at 1.2% strain) observed in experiments as shown in **Figure 4c**. This indicates that tensile strain primarily modifies the spin-phonon relaxation rate in the spin-down channel. The strain dependence of the spin-down demagnetization rate is dominant and drives the overall demagnetization rate of the system. This is consistent with the experimental observation that tensile strain enhances and accelerates demagnetization in $Fe_3GaTe_2$.

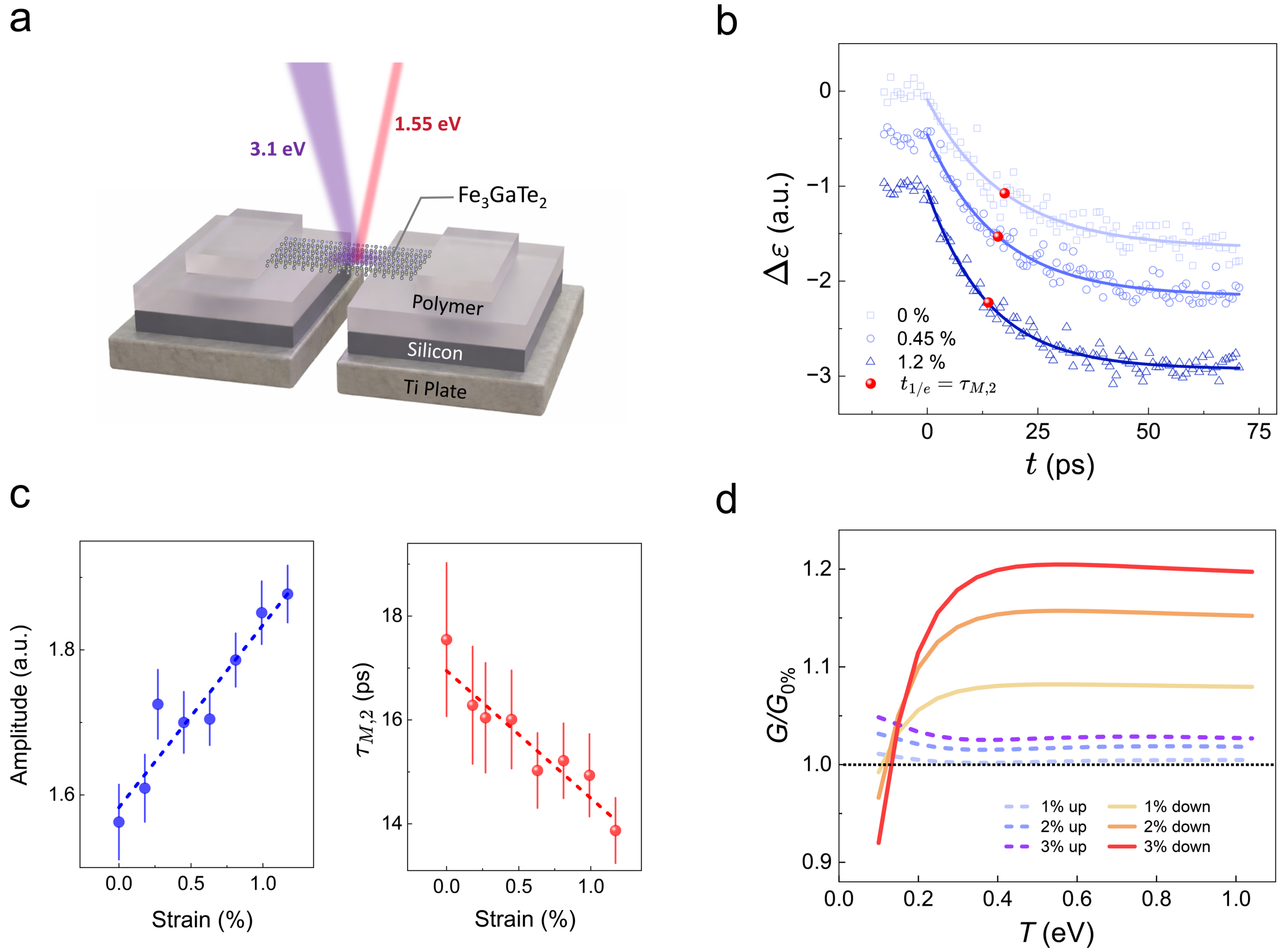

**Figure 4: In situ strain control of ultrafast demagnetization in $Fe_3GaTe_2$. a,** Schematic illustration of the strain dependent TR-MOKE measurement. An exfoliated $Fe_3GaTe_2$ flake is transferred onto a PCL coated gapped Si substrate mounted on a Ti plate. The gap is around 20 µm. The strain sample is excited by a 400 nm pump pulse and probed by an 800 nm pulse. **b,** Transient pump-induced Kerr ellipticity change, $\Delta\varepsilon$, measured at selected tensile strains under an out-of-plane magnetic field of 1800 Oe and a pump fluence of 1 mJ/cm$^2$. Curves are vertically offset for clarity. Symbols represent experimental data, and solid lines denote fits using the slow component of Eq. (1). Red dots indicate the characteristic $1/e$ demagnetization point. **c,** Strain dependence of the slow-component amplitude $A_2$ and corresponding demagnetization time $\tau_{M,2}$, extracted from the data in **b**. **d,** Electron-phonon relaxation rate ratio relative

to 0% strain as a function of effective electron temperature under uniaxial strain 1%, 2% and 3%. for the spin-up and spin-down channel.

## Conclusion

In summary, we demonstrate that uniaxial tensile strain provides an effective in situ route to tune both the equilibrium magnetic response and ultrafast spin dynamics of the room-temperature van der Waals ferromagnet $Fe_3GaTe_2$. Static RMCD measurements reveal a pronounced strain-induced enhancement of the coercive field and remanent magneto-optical response, consistent with strengthened effective perpendicular magnetic anisotropy. Time-resolved MOKE measurements further show that tensile strain accelerates the intrinsic picosecond demagnetization dynamics, establishing mechanical deformation as a means of tuning nonequilibrium spin relaxation. Remarkably, strain accesses a fast-relaxation regime under weak optical excitation, demonstrating a control mechanism distinct from conventional fluence-driven demagnetization in the unstrained material. This distinction indicates that strain does not simply mimic stronger optical excitation, but provides an independent route to tune the underlying spin-relaxation process while reducing the required optical energy. Combined with first-principles calculations, our results suggest that lattice distortion modifies the spin-lattice energy transfer underlying the ultrafast response. More broadly, these findings establish mechanical strain as a general strategy for on-demand control of ultrafast magnetic dynamics by tuning the magnetic free-energy landscape and the associated lattice-mediated spin-relaxation channels. The mechanical flexibility and strong magnetoelastic sensitivity of van der Waals magnets therefore make them a promising platform for mechanically programmable high-speed spintronic functionalities.

## Methods

### Single crystal synthesis:

$Fe_3GaTe_2$ single crystals were grown with the chemical vapor transport (CVT) method. 1 g of stoichiometric amount (3:1:2) of iron powder (99.9%), gallium ingot (99.999%), and tellurium chunk (99.9999%) are mixed with 70 mg of iodine lump (99.999%), loaded into a quartz ampule and sealed under vacuum ($10^{-5}$ Torr). Subsequently, the sealed ampule was placed into a two-zone furnace and heated under a temperature gradient with the source material at the hot end. The hot (cold) end was ramped to 750 °C (725 °C) over a 24 h period and held at this temperature gradient for two weeks. Finally, the sample naturally cooled down by turning the furnace off. The process yielded shiny and flat crystals.

### Strain device fabrication

To prepare the substrate, a trench with customized width is formed by partially etching (~50 μm) a 100-μm-thick Si chip, followed by surface functionalization with Polycaprolactone (PCL) film and substrate mounting on the strain cell (Razorbill CS120). The substrate fabrication, functionalization and mounting procedure followed our previous report[38]. The $Fe_3GaTe_2$ flakes are freshly exfoliated on a PDMS cube and identified with optical microscopy. Afterwards, the PDMS/sample stack was aligned over the trench in the Si chip. After making contact between PDMS/sample and substrate, the entire transfer stack was heated to and held at 60 °C (above the glass transition temperature of PCL), for 20 s. Finally, the entire stack was cooled to room temperature and the PDMS was released from the sample/substrate.

### Reflective magnetic circular dichroism measurements

Reflective magnetic circular dichroism (RMCD) measurements were carried out using a 632.8 nm He-Ne laser and a Quantum Design OptiCool cryostat. A 20x objective focused the probe beam to a ~2 μm diameter, with the optical power kept under 3 μW. The incident beam was intensity-modulated at 773 Hz using a mechanical chopper, and its polarization was alternated between right- and left-circular states using a photoelastic modulator (PEM, Hinds Instruments) modulated at $f$ = 50 kHz with a $\pi$ retardation, followed by a quarter-wave plate. A two-channel lock-in amplifier (OD1022D) recorded the reflected signal, capturing the total reflectance ($R$) at the chopper frequency and the differential reflectance ($\Delta R$) at the PEM's second harmonic ($2f$). The final RMCD signal is defined as $\Delta R/R$.

**Time-resolved magneto-optical Kerr effect measurements**

Time-resolved magneto-optical Kerr effect (TR-MOKE) measurements were performed using a Ti:sapphire femtosecond amplifier (Astrella, 40 fs, 800 nm and 1 kHz repetition rate). The laser output was split into pump and probe branches. The pump was frequency-doubled to 400 nm in a barium borate (BBO) crystal (Eksma Optics), whereas the 800 nm probe was amplitude-modulated at 491 Hz by a mechanical chopper. Both beams were focused onto the sample at near-normal incidence with a 1-inch lens, producing spot sizes of ~60 μm for the pump and ~10 μm for the probe. The reflected probe was analyzed by a quarter-wave plate and a Wollaston prism, and the orthogonal polarization components were detected with balanced silicon photodiodes. The differential signal ($\Delta I$) was measured using a lock-in amplifier referenced to the chopper frequency, and the pump-induced Kerr ellipticity change was defined as $\Delta\eta = \Delta I/2I$, where $I$ is the total reflected probe intensity of each measurement. The external magnetic field was applied using a pair of permanent magnets mounted on a translation stage, allowing the field strength to be tuned by adjusting the stage's displacement (see Supplementary Section S6).

**First-principles calculations**

Density functional theory calculations are performed using the Vienna Ab Initio Simulation Package (VASP)[55]. The projector augmented wave (PAW) method[56,57] implemented in VASP is used with a kinetic energy cutoff of 500 eV. The local density approximation (LDA) is used as the exchange-correlation functional. Structures are fully relaxed until the force on each atom is less than 0.01 eV/Å. The relaxed lattice constants of the primitive cell are *a*=*b*=3.944 Å and *c*=15.822 Å, which are in good agreement with experimental ones *a*=3.986 Å and *c*=16.229 Å[58]. To simulate a uniaxial strain, a rectangular $1 \times \sqrt{2} \times 1$ unit cell is used. A uniaxial strain is defined as $\varepsilon = (a - a_0)/a_0$, where $a_0$ is the lattice constant at equilibrium. The lattice constants under uniaxial strain are shown in **Figure S15**. The calculated magnetic moment of Fe is 1.99 $\mu_B$ and 1.33 $\mu_B$ for each kind of Fe with an averaged magnetic moment of 1.77 $\mu_B$, which is in excellent agreement with experimental magnetic moment of 1.68 $\mu_B$ at 3 K[58]. Magnetic anisotropy energy (MAE) is calculated as the total energy difference between in-plane and out-of-plane magnetization with spin-orbit coupling turning on. The k-point convergence is tested, and a $33 \times 15 \times 3$ k-point mesh gives a converged MAE as shown in **Figure S14**. The calculated MAE is $9.63 \times 10^5$ J/m$^3$ which is comparable to the experimental result of $4.79 \times 10^5$ J/m$^3$ at 300 K[58].

**Acknowledgements**

We thank Prof. Mark Rzchowski at UW-Madison for providing the scanning permanent magnet setup used in our TR-MOKE measurements. We also thank Dr. Chen Liu at KAUST for helpful discussions. This research was primarily supported by NSF through the University of Wisconsin Materials Research Science and Engineering Center (DMR-2309000).

**Author Contributions**

F.F., D.A.R., and J.X. conceived the research. Y.P., D.A.R, and J.X. supervised the project. Y.H. and J.K. synthesized the bulk $Fe_3GaTe_2$ crystal under the guidance of D.A.R. The unstrained and strained samples were prepared and calibrated by Y.H. and F.F. The AFM measurements were performed by R.B. with assistance from F.F. The SQUID measurement was conducted by J.K. F.F. performed the static RMCD and TR-MOKE measurements and analyzed the data with J.X. The first-principles calculations were performed by W.F. and J.S. under the guidance of Y.P. All authors discussed the results and jointly wrote the manuscript.

**Conflict of interest**

The authors declare no competing interests.

**Data availability**

The data that support the findings of this study are available from the corresponding authors upon reasonable request.